\documentclass[conference,letterpaper]{IEEEtran}
\IEEEoverridecommandlockouts

\usepackage{cite}
\usepackage{amsmath,amssymb,amsfonts}
\usepackage{algorithmic}
\usepackage{graphicx}
\usepackage{textcomp}
\usepackage{xcolor}
\usepackage{tabularx}
\usepackage{array}
\def\BibTeX{{\rm B\kern-.05em{\sc i\kern-.025em b}\kern-.08em
    T\kern-.1667em\lower.7ex\hbox{E}\kern-.125emX}}
\begin{document}

\title{The Balancing Act of Policies in Developing Machine Learning Explanations\\}

\author{\IEEEauthorblockN{Jacob Tjaden}
\textit{Colby College}\\
Waterville, Maine U.S.A.\\}

\maketitle

\section{Introduction}
Due to the nature of opaque machine learning (ML) models, software engineers and data scientists struggle to understand how ML models make decisions \cite{rudin2019stop}. Explainability research aims to provide transparency for these models \cite{lipton2018mythos} through two types of explanations. Global explanations describe how a model works generally and provide insight into its accuracy, biases, and fairness. Local explanations describe individual predictions made by the model in specific use cases.

Explanations of ML models hold intrinsic value in improving transparency \cite{colaner2022explainable}. In the medical landscape, providing explanations for patients can encourage trust in the model \cite{lipton2018mythos}. Medical professionals can use explanations to determine reliability in model diagnoses \cite{bhavsar2021medical}. Data scientists can use explanations to debug, test, and evaluate the model \cite{krishnan2017palm}. 

Despite their value, however, producing explanations that are meaningful to a variety of stakeholders presents significant challenges \cite{cai2019hello}. The internal design of opaque ML models makes it difficult to translate model predictions into intelligible explanations \cite{rudin2019stop}. Stakeholders have varying priorities and background experiences, which makes the development of widely accessible explanations challenging \cite{nahar2024regulating}. Additionally, as shown in Figure 1, there are often notable technical limitations of instruments that create explanations \cite{rudin2019stop}.

\begin{figure}[h]
  \centering
  \vspace{-5pt}
  \includegraphics[width=0.45\textwidth]{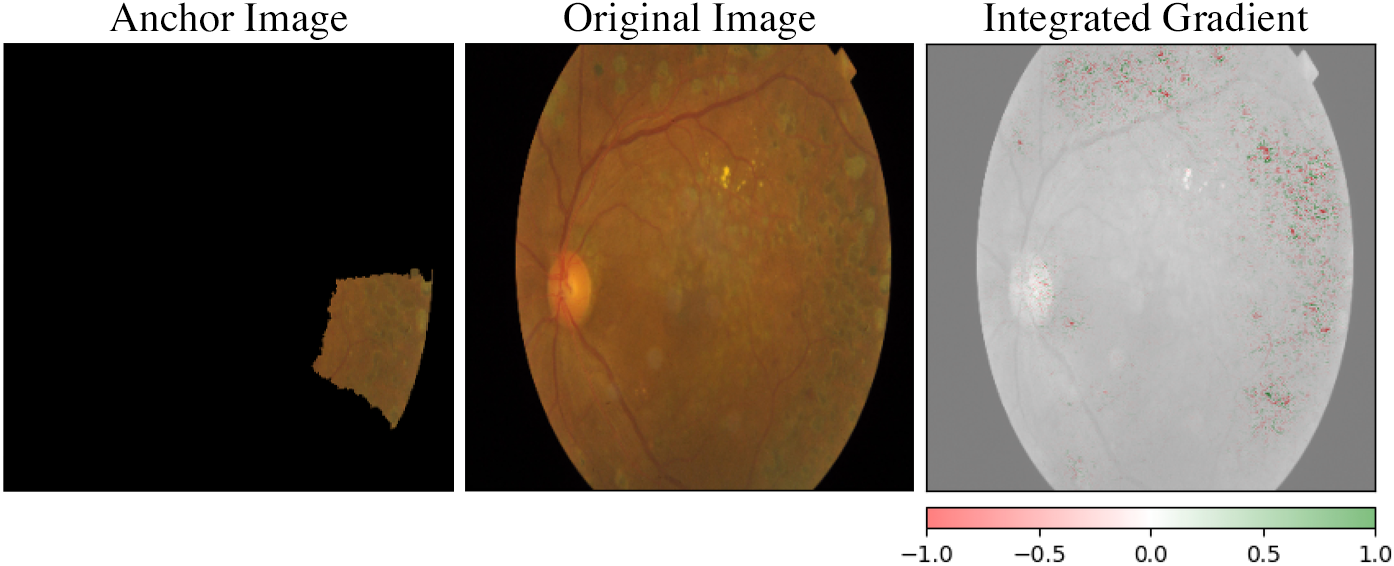}
  \vspace{-5pt}
  \caption{Local Explanation Examples: The Anchor image is meaningless to stakeholders, while the integrated gradient is more accessible to trained users.}
  \vspace{-5pt}
\end{figure}

To address these challenges, policies can be used to inform explanations as guidelines and requirements for providing transparency. Policies can take the form of government regulation, such as the European Union AI Act and White House executive order on the trustworthy development of AI, or in-house for companies that strive to promote responsible AI practices. However, developing effective policies is difficult, and there is little work in the field exploring what form of policies best yields changes in developer behavior \cite{nahar2024regulating}. Our research explores the relationship between policies and explanations. We strive to answer the question: \textit{how do data scientists interpret policies, react to different policy purposes, and provide evidence for compliance?}

We performed a large-scale controlled experiment with 124 participants and six policy conditions, and analyzed the effect of each condition on the implementation of explanations by developers. Our findings suggest that while the length of policies influences engagement with certain policy requirements, the policy's purpose has no measurable impact, and the overall quality of explanations is poor. This highlights the difficulty of creating effective policies, which may benefit from greater emphasis on engaging with stakeholder perspectives.

\begin{table*}[t]
\centering
\vspace{-10pt}
\caption{Examples of Policy Requirements}
  \vspace{-5pt}
\begin{tabular}{ | m{0.9in} | m{5.8in} | }
  \hline
  General Reqs. & \scriptsize \texttt{Designers, developers, and deployers of automated systems should provide generally accessible plain language documentation including clear descriptions of the overall system functioning and the role automation plays, notice that such systems are in use, the individual or organization responsible for the system, and explanations of outcomes that are clear, timely, and accessible.} \\ 
  \hline
  Specific Req. \#3 & \scriptsize \texttt{Describe how the automation (model) works generally. Provide evidence that the documentation is effective for the policy purpose. \textit{[Guidance: Where possible identify general mechanisms or factors that most strongly influence the automation.]}} \\ 
  \hline
\end{tabular}
\vspace{-10pt}
\end{table*}

\section{Methods}

We conducted an IRB-approved classroom experiment where 124 students were each provided an explanation policy and an ML-powered screening device \cite{abramoff2018pivotal}. Participants were tasked with developing explanations for this opaque ML-powered device, which detects diabetic retinopathy (DR) using smartphone images. The screening device assesses DR severity on a scale of 0 to 4 using images of patients' eyes, with the age and gender of patients as additional inputs.

Participants were randomly assigned one of six policies varying in both length and purpose. As shown in Table 1, policies could be short, containing a paragraph description of general requirements for explanations, or long, with an explicit list of eight specific requirements in addition to the general requirements. Each policy also contained one of three purposes: \textit{1) to preserve the dignity of individuals}, \textit{2) to enable effective human-AI collaboration}, or \textit{3) no purpose}.  Participants submitted one global explanation and two local explanations, and provided a self-evaluation document analyzing their compliance with their assigned policy.

We used standard qualitative methods \cite{lazar2017research} to analyze submissions. We first conducted open coding~\cite{saldana2021coding} on seventeen submissions and developed fifty codes. Next, we conducted axial coding to generate a codebook, followed by closed coding~\cite {saldana2021coding} and several refinements of the codebook. Our codebook covers two categories: a participant self-assessment of compliance with eight policy requirements, and an external assessment by the researcher of four requirements in addition to the eight self-assessed. For each requirement, we considered whether participants claimed to comply with the requirement, explicitly stated they did not comply, or did not discuss their compliance. Finally, we identified trends in explanations.

\section{Results}

\subsection{General Trends}

Participant explanations generally did not comply with policy requirements, and participants tended to claim that they complied with requirements even if they did not. Across local explanations, 85\% of participants used Anchor or LIME libraries—tools that highlight the image regions most influential in the model's DR prediction. 19\% of participants made use of libraries such as SHAP that provide the importance of additional variables used by the model, such as age and gender. 

Despite the commonality of these techniques, the resulting explanations often appeared to be meaningless to stakeholders, and their practical benefits were unclear. The prevalence of Anchor and LIME libraries may reflect participants' exposure prior to the study, but also suggests that developers may rely on familiar explanation methods instead of optimal alternatives.

The global explanations can be grouped into three main categories. Results from \textit{accuracy testing} of the model were provided by 81\% of participants. 50\% of participants used explanations related to \textit{fairness or subgroup testing} that showed model performance by age or gender. 76\% of global explanations contained \textit{diagnosis data analysis}, which measured the model's prediction accuracy for each diagnosis severity. This method generally showed trends and biases of the model's prediction diagnosis compared to the actual diagnosis, but most explanations provided little more than prediction accuracy.

\subsection{Specific Findings}

\textbf{Policy Purpose.} 	The three policy purposes did not have a significant effect on participant responses. Regardless of purpose, explanations often failed to consider stakeholder priorities, and we were generally unable to determine a participant's purpose from their explanations.

\textbf{Policy Length.} 	Participants with longer policies were more likely to engage with specific requirements, while short policy participants were slightly better at complying with general requirements. Figure 2 captures this difference in participant self-assessment. 

\begin{figure}[h]
  \centering
  \includegraphics[width=\linewidth]{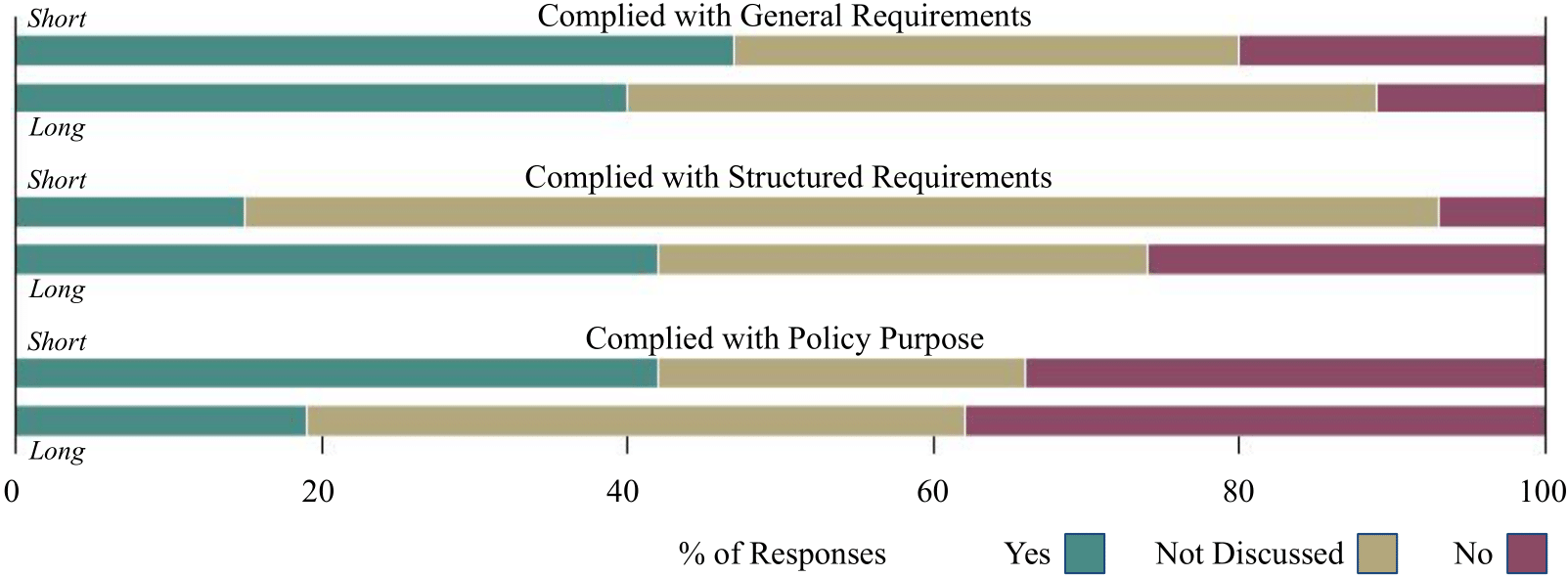}
  \vspace{-15pt}
  \caption{Participant Self-Evaluated Compliance by Policy Length}
\end{figure}

\textbf{Unexpected Findings.}	As an exception to these trends, short policy participants were more likely to \textit{provide evidence that their documentation was effective for their policy purpose}, despite this requirement only appearing in the long policy. 58\% of developers with the short policy claimed that their explanation was effective for the policy purpose, whereas only 28\% of developers with the long policy claimed effectiveness.

There are two potential reasons for this difference. One likely contributor is the freedom granted in shorter policies, where the policy purpose may more easily manifest in explanations. Alternatively, the policy purpose may become diluted within the long policy's requirements, whereas in the short policy the purpose may seem more meaningful to developers.

\section{Contributions}

Our study highlights the relationship between policies and ML transparency. While longer, structured policies result in increased developer engagement with specific requirements, shorter policies offer flexibility that can improve alignment with policy purposes. However, the overall poor quality of explanations overshadows this difference. The explanations provided were largely ineffective and rarely engaged with stakeholder perspectives, despite clear policy guidance. This emphasizes the challenge of developing effective policies, and the importance of stakeholder consideration in producing meaningful ML explanations. Future work could explore stakeholder-oriented policies, where each policy requirement generates benefit for a targeted stakeholder.

\bibliographystyle{ieeetr}
\bibliography{references}

\begin{thebibliography}{10}

\bibitem{rudin2019stop}
C.~Rudin, ``Stop explaining black box machine learning models for high stakes decisions and use interpretable models instead,'' {\em Nature machine intelligence}, vol.~1, no.~5, pp.~206--215, 2019.

\bibitem{lipton2018mythos}
Z.~C. Lipton, ``The mythos of model interpretability: In machine learning, the concept of interpretability is both important and slippery.,'' {\em Queue}, vol.~16, no.~3, pp.~31--57, 2018.

\bibitem{colaner2022explainable}
N.~Colaner, ``Is explainable artificial intelligence intrinsically valuable?,'' {\em AI \& society}, pp.~1--8, 2022.

\bibitem{bhavsar2021medical}
K.~A. Bhavsar {\em et~al.}, ``Medical diagnosis using machine learning: a statistical review,'' vol.~67, no.~1, pp.~107--125, 2021.

\bibitem{krishnan2017palm}
S.~Krishnan and E.~Wu, ``Palm: Machine learning explanations for iterative debugging,'' in {\em Proc. 2nd workshop on human-in-the-loop data analytics}, pp.~1--6, 2017.

\bibitem{cai2019hello}
C.~J. Cai {\em et~al.}, ``"hello ai": uncovering the onboarding needs of medical practitioners for human-ai collaborative decision-making,'' {\em Proc. ACM on Human-computer Interaction}, vol.~3, no.~CSCW, pp.~1--24, 2019.

\bibitem{nahar2024regulating}
N.~Nahar {\em et~al.}, ``Regulating explainability in machine learning applications--observations from a policy design experiment,'' in {\em Proc. Conf. ACM FAccT}, pp.~2101--2112, 2024.

\bibitem{abramoff2018pivotal}
M.~D. Abr{\`a}moff {\em et~al.}, ``Pivotal trial of an autonomous ai-based diagnostic system for detection of diabetic retinopathy in primary care offices,'' {\em NPJ digital medicine}, vol.~1, no.~1, p.~39, 2018.

\bibitem{lazar2017research}
J.~Lazar, J.~H. Feng, and H.~Hochheiser, {\em Research methods in human-computer interaction}.
\newblock Morgan Kaufmann, 2017.

\bibitem{saldana2021coding}
J.~Salda{\~n}a, ``Coding techniques for quantitative and mixed data,'' {\em Routledge reviewer’s guide to mixed methods analysis}, pp.~151--160, 2021.

\end{thebibliography}

\end{document}